\documentclass[]{sn-jnl}
\usepackage{graphicx}%
\usepackage{multirow}%
\usepackage{amsmath,amssymb,amsfonts}%
\usepackage{amsthm}%
\usepackage{mathrsfs}%
\usepackage[title]{appendix}%
\usepackage{xcolor}%
\usepackage{textcomp}%
\usepackage{manyfoot}%
\usepackage{booktabs}%
\usepackage{algorithm}%
\usepackage{algorithmicx}%
\usepackage{algpseudocode}%
\usepackage{listings}%

\usepackage[numbers]{natbib}
\begin{document}

\title[A latent class approach to assess the effects of dynamic adherence to polytherapy in heart failure patients]{A latent class approach to assess the effects of dynamic adherence to polytherapy in heart failure patients}

\author*[1,2]{\fnm{Nicole} \sur{Fontana}}\email{nicole.fontana@polimi.it}
\equalcont{These authors contributed equally to this work.}

\author[2]{\fnm{Laura} \sur{Savaré}}
\equalcont{These authors contributed equally to this work.}



\author[2,3-5]{\fnm{Emanuele} \sur{Di Angeloantonio}}

\author[1,2]{\fnm{Francesca} \sur{Ieva}}

\affil[1]{\orgdiv{MOX, Department of Mathematics}, \orgname{Politecnico di Milano}, \orgaddress{\state{Italy}, \country{Milan}}}

\affil[2]{\orgdiv{HDS, Health Data Science Center}, \orgname{Human Technopole}, \orgaddress{\state{Italy}, \country{Milan}}}



\affil[3]{\orgdiv{Blood and Transplant Research Unit in Donor Health and Behaviour}, \orgname{National Institute for Health and Care Research}, \orgaddress{\state{UK}, \country{Cambridge}}}

\affil[4]{\orgdiv{Dept of Public Health \& Primary Care}, \orgname{University of Cambridge}, \orgaddress{\state{UK}, \country{Cambridge}}}

\affil[5]{\orgdiv{British Heart Foundation Centre of Research Excellence}, \orgname{University of Cambridge}, \orgaddress{\state{UK}, \country{Cambridge}}}

\affil[6]{\orgdiv{Health Data Research UK Cambridge}, \orgname{Wellcome Genome Campus and University of Cambridge}, \orgaddress{\state{UK}, \country{Cambridge}}}


\abstract{Heart failure (HF) treatment relies heavily on pharmacotherapy, particularly combining multiple therapies as recommended by clinical guidelines. However, non-adherence to prescribed regimens remains a significant challenge, contributing to increased hospitalizations and poorer patient outcomes. This study introduces a novel methodological pipeline that integrates Latent Markov Models (LMM) with dynamic adherence modeling to evaluate adherence behaviors and their impact on HF rehospitalization.

Using administrative healthcare data from Lombardy, Italy, we analyzed 6,818 patients hospitalized for HF between July and December 2020. Adherence was assessed monthly over a six-month observation period, and adherence profiles were linked to clinical outcomes using Cox regression. Seven latent behavioral profiles were identified, reflecting varying levels and trajectories of adherence.

The findings revealed that higher adherence levels significantly reduced the risk of rehospitalization. Patients with consistently high adherence exhibited a 56\% lower risk of HF rehospitalization compared to those with low adherence. Importantly, improving adherence during the observation period was associated with better survival probabilities, highlighting the potential benefits of timely interventions. Additionally, adherence behaviors were influenced by factors such as age, comorbidity burden, and hospitalization during the observation period.

This study underscores the importance of dynamic and personalized strategies to monitor and enhance adherence to polytherapy. By linking adherence patterns to clinical outcomes, the proposed approach offers actionable insights for improving patient management and reducing the burden of HF on healthcare systems.}

\keywords{Latent Markov model, Polytherapy, Dynamic adherence, Heart failure, Administrative database}

\maketitle

\section{Introduction}\label{sec1}\label{CH1}

Heart failure (HF) poses a significant global health challenge, contributing substantially to morbidity, mortality, and healthcare burden worldwide~\citep{cite_1}. Despite advances in pharmacological treatments, HF remains associated with high rates of hospitalization and adverse outcomes. Guidelines recommend the use of combination therapies, which have been shown to improve survival and reduce morbidity~\citep{cite_4}. Among these, angiotensin-converting enzyme inhibitors (ACE-I), angiotensin-receptor blockers (ARBs), beta-blockers (BB), and mineralocorticoid receptor antagonists (MRA) form the backbone of HF pharmacotherapy~\citep{cite_2, cite_3}. However, the effectiveness of these therapies is often undermined by the pervasive issue of medication non-adherence~\citep{cite_5}.

Non-adherence to prescribed therapies is common among HF patients, with rates ranging widely from 10\% to 93\% depending on the measurement approach~\citep{cite_6, cite_7}. This behavior is strongly linked to worse health outcomes, including increased hospitalizations and higher healthcare costs~\citep{cite_5, cite_8}. Conversely, higher adherence rates have been consistently associated with a significant reduction in HF-related hospital readmissions~\citep{cite_21}, emphasizing the critical need to promote adherence to prescribed therapies. Yet, understanding and improving adherence remains challenging, as adherence behaviors vary over time due to factors such as patient engagement, clinical visits, and disease progression.

Improving medication adherence is crucial to enhancing health outcomes and reducing the burden on healthcare systems. Over the years, progress in understanding HF pathophysiology has led to improvements in treatment~\citep{cite_10}. However, this evolving landscape highlights the need for systematic approaches that evaluate adherence to polytherapy while accounting for the growing complexity of pharmacological options. Moreover, the COVID-19 pandemic has introduced additional challenges to healthcare delivery, disrupting medication adherence, follow-up visits, and timely hospitalizations, thereby potentially impacting HF management~\citep{cite_covid1, cite_covid2}. Understanding adherence patterns during this period is essential for optimizing care in similar scenarios and comparing effectiveness of the management of care during emergencies and routinary periods.

Moreover, given the importance of combining therapies for HF treatment, there is a need for tools capable of describing adherence to multiple drugs across diverse patient populations. Prior research on adherence to polytherapy~\citep{cite_11, cite_12, cite_13} has often been limited to ``user-only" cohorts~\citep{cite_14}, overlooking the heterogeneity in prescribed regimens. Doctors’ prescriptions vary based on HF severity, underlying causes, and patient comorbidities~\citep{cite_4}. Consequently, it is essential to employ methods that evaluate adherence dynamically and capture its influence on clinical outcomes.
Therefore, despite the guidelines suggesting mainly the use of the RAS, BB and MRA drugs, doctors’ prescriptions can be different from them based on the underlying cause of the heart failure, the severity of the condition, intolerance, and other medical conditions patients may have~\citep{cite_4}. We are therefore interested in developing a method that not only evaluates adherence to polytherapy but can also evaluate it on a cohort of patients with different combinations of prescribed drugs. At the same time, it is interesting to catch the effects of the dynamics of drug adherence on patients’ outcomes since, during the course of therapy, drug intake and adherence patterns can change. Studies indicate that adherence is higher after the initial diagnosis of the disease and then tends to decline over time~\citep{cite_15}, and it has also been observed that patients typically improve their medication adherence just before and after a medical visit~\citep{cite_16}. Initially and still today, in some studies, medication adherence is modelled using binary variables without taking into account changes over time, which results in a lack of fundamental information~\citep{cite_17}. In the available literature and current practice, several methods were devel- oped to model this measure as a time-varying variable~\citep{cite_18,bijlsma2016estimating}. This approach allows researchers to account for changes in adherence over time and can be an effective and accurate way to represent drug consumption. To address this, we propose a novel methodological pipeline that integrates Latent Markov Models (LMM)~\citep{cite_20} with dynamic adherence modeling to analyze categorical longitudinal data on drug adherence derived from administrative databases and evaluate their impact on HF rehospitalization.
Our approach investigates the dynamic nature of adherence to multiple therapies, uncovering insights into evolving patterns and linking them to clinical outcomes. By integrating adherence findings into a prognostic model, we aim to assess their impact on HF rehospitalization. This innovative pipeline highlights the importance of continuous adherence monitoring and underscores the need for interventions to support medication adherence over time.

The remaining sections of this paper are organized as follows. Section~\ref{CH2} describes the data extraction, inclusion criteria, and representation of pharmacological covariates. Section~\ref{CH3} outlines the statistical methodologies. Key results are presented in Section~\ref{CH4}, followed by a discussion of strengths and limitations in Section~\ref{CH5}.

\section{Dataset}\label{CH2}
\subsection{Study setting and data sources}
Data from the healthcare utilization databases of Lombardy, a northern region of Italy accounting for 10 million residents, about patients hospitalized for heart failure between July 1, 2020 and December 31, 2020 are analyzed. The first six months of 2020 were excluded to avoid introducing the potential effects of the first COVID-19 lockdown, which was the most restrictive and likely influenced drug purchasing habits as well as access to hospitals.
Information from the different databases is linked via a unique anonymous identification code to protect individual privacy. The data linkage spans the period from January 1, 2012 to July 1, 2023 to provide a longitudinal view of patient interactions across different healthcare settings. Hospitalization records provide information on primary diagnosis, coexisting conditions, and procedures coded using the International Classification of Diseases, 9th Revision Clinical Modification (ICD-CM-9) classification system~\citep{cite_21}. Drug purchase records code each drug according to the Anatomical Therapeutic Chemical (ATC) classification system~\citep{cite_22} and information about the number of days of treatment covered by it is almost always extracted using the number of tablets and the Defined Daily Dose (DDD) metric~\citep{cite_23}. However, because BB and MRA for heart failure are likely to be prescribed at lower dosages than those established for treating hypertension~\citep{cite_24,cite_25}, the corresponding dosage was adjusted following~\citep{cite_26} for BB and by a working group of experienced clinicians for MRA. Information about drug purchases is used for computing adherence to medication, as commonly done with administrative data~\citep{corrao_adherence}.

\subsection{Study population and cohort selection}
This study analyzed a cohort of 6,818 patients who underwent their first hospitalization for heart failure, referred to as the index hospitalization, between July 1, 2020, and December 31, 2020. The initial dataset comprised 18,575 patients hospitalized for HF within this period. However, several exclusion criteria were applied to ensure a homogenous and relevant population. Patients younger than 50 years of age (\textit{n} = 342) and those who were not beneficiaries of the regional health service in the three years preceding the index hospitalization (\textit{n} = 144) were excluded. To capture only incident cases of HF, a three-year wash-out period before the index hospitalization was employed, leading to the exclusion of 6,280 patients with a prior diagnosis of HF. Additionally, 8,624 patients who did not initiate pharmacological treatment with RAS, BB, or MRA within three months of discharge were removed. Finally, 884 patients with less than six months of follow-up after the index hospitalization were excluded. After applying these criteria, the final cohort consisted of 6,818 patients.
Comorbidities extracted from the ICD-9-CM codes of each hospitalization, combined with drug purchases, were used to compute the Multisource Comorbidity Score (MCS)~\citep{cite_27}. This index combines hospital diagnoses and drug prescriptions to provide a tool capable of measuring the patient's overall comorbidity burden. Detailed information about the cohort selection are available in Figure S1. HF diagnoses were coded using the ICD-9 system, and medications were classified based on the ATC coding system (Table S1).

At the index hospitalization, the patients in the cohort had a mean age of 78.18 years, and 47.8\% were women. The median MCS, which combines hospital diagnoses and drug prescriptions to measure the overall comorbidity burden, was 10 (IQR 7–15). From the index hospitalization, patients were followed for a median duration of 720 days (IQR 530–720), during which 1,553 patients (22.8\%) experienced rehospitalization for HF. Regarding drug purchases, BB were the most commonly purchased medications, with 88.5\% (\textit{n} = 5,750) of patients purchasing at least one. This was followed by RAS inhibitors, purchased by 74.4\% (\textit{n} = 4,814) of patients, and MRAs, which had the lowest purchase rate at 53.2\% (\textit{n} = 3,444). Descriptive analysis of the patients is available in Table S2. During the observation period, 494 patients (7.4\%) experienced a hospitalization related to COVID-19. This significant external factor was considered in the analyses, and sensitivity tests were conducted to account for its potential impact on drug adherence and clinical outcomes.

\subsection{Study design} The study was structured into two distinct phases. The first phase, the observation period, spanned six months from the discharge date of the index hospitalization. During this phase, adherence to pharmacological therapy was assessed, focusing on RAS inhibitors, BBs, and MRAs. This was followed by a follow-up period, which began immediately after the observation period and continued until the occurrence of an HF rehospitalization, death, loss to follow-up, or the study endpoint on July 1, 2023. The follow-up phase included only patients who were alive and under observation at the conclusion of the initial six months, with the primary outcome being time to heart failure rehospitalization. To ensure the robustness of the findings, sensitivity analyses were conducted by varying the lengths of both the observation and follow-up periods (for further details, see Section 3.5).

\subsection{Adherence measure to a single drug}
To effectively measure and analyze drug adherence over time, we computed the cumulative time-varying adherence to each drug, a time-dependent version of the  Proportion of Days Covered (PDC) measure~\citep{cite_30}. This method, described comprehensively in \citep{cite_29}, allows for dynamic assessment of patient adherence patterns.
Cumulative time-varying adherence is conceptualized as a three-level, time-dependent categorical variable that captures the adherence levels of the $i$-th patient to a specific drug at any given time $t$. 
Categorizing adherence into distinct levels allows for a clearer understanding and interpretation of adherence patterns and their potential impact on clinical outcomes, providing clinicians with actionable insights into patient behavior and allowing for targeted interventions. This approach is detailed across multiple sources \citep{new_1, new_2, new_3, new_4, new_5} and provides a robust framework for monitoring adherence over the course of treatment.
\begin{equation}
         \label{eq1.1}
        \text{adherence}_i^{(t)} = \begin{cases}
            0, &\text{if } \frac{\text{com\_month}_i(t)}{t} \in \text{[0, 0.25)},\\
            1, &\text{if } \frac{\text{com\_month}_i(t)}{t} \in \text{[0.25, 0.8),}\\
            2, &\text{if } \frac{\text{com\_month}_i(t)}{t} \in \text{[0.8, 1].}\\
            \end{cases}
\end{equation}
The variable $\text{com\_month}_i(t)$ in Equation~(\ref{eq1.1}) was defined for each patient $i$ as the cumulative count of distinct coverage months up to month $t$. During its computation, we considered the number of distinct days covered by the drug purchase; this means that in case of overlapping periods, it was considered the first purchase entirely and only the days of the second one not covered by the first. Adherence was calculated monthly for the entire observation period for RAS, BB and MRA only for those patients who are users of the considered drugs. To be considered a user of a particular drug, a patient must have purchased it at least once during the observation period. A crucial aspect of our analysis is calculating adherence only for users. This approach is more effective than the one often adopted in similar studies as it allows us to manage information about non-users during the modeling phase. In fact, we can differentiate between non-users and non-adherent, which provides a clearer understanding of drug use patterns and avoids any potential biases that may arise from including data from non-users as non-adherent information. 
In pharmacoepidemiologic studies based on the Lombardy dataset, patients are often classified as having low adherence to treatment if their PDC is below 25\%, and as highly adherent if it exceeds 75\%. These thresholds were established based on previous studies using this dataset, which demonstrated a clear association between adherence levels and cardiovascular outcomes or mortality. Specifically, patients in the highly adherent group exhibited significantly lower rates of adverse outcomes \citep{new_1, new_2}.
In alternative settings, such as in this study, a more conventional threshold of PDC greater than 80\% is used to define optimal adherence \citep{new_5}. Regardless of the specific threshold applied, these classifications are evaluated through sensitivity analyses, which test the robustness of the results by varying the adherence thresholds.

\section{Methodology}\label{CH3}
In this section, we outline our methodological pipeline for applying the latent Markov model in pharmacoepidemiology. We begin with a comprehensive exploration and definition of these models, emphasizing how this approach uncovers underlying patterns in pharmaceutical drug utilization. Next, we demonstrate how to integrate the results from the latent Markov model into a prognostic model to assess the relationship between these patterns and clinical outcomes. Finally, we conduct sensitivity analyzes to assess the robustness of our findings.

\subsection{Latent Markov model on pharmacoepidemiology setting}
\label{3.1}
Latent Markov (LM) models represent an important class of latent variable models used to analyze longitudinal data~\citep{dlv_bartolucci}. These models assume the existence of a latent process of interest that is not directly observable and affects the distribution of the observed variables~\citep{cite_20}. When applied to the longitudinal assessment of drug adherence, the LM approach hinges on the presence of a latent process, representing the patient's varying levels of \textit{willingness} to take the medications. Where, with \textit{willingness}, we mean the individual tendency to take drugs or to follow the medical indication about not take it, so on being \textit{non-user} of the drugs.
Two primary motivations drive the utilization of LM models in quantifying overall drug adherence. First, account for measurement errors in the computed adherence variables. Secondly, identify different patients' behaviours, referred to as latent states, within the broader patient cohort, along with tracking their temporal changes.

A LM model can be seen as a latent class model~\citep{cite_33}, where patients can change latent states during the observed period. These model are an extension of the Markov chain model, which account for measurement errors~\citep{cite_20}. In particular, the latent variable is observed with a measurement error of the response variable. 
LM models are based on the \textit{local independence} assumption, which implies that the response variables are conditionally independent given the latent variables since the latent variables represent the unique explanatory factor of the outcomes. When it comes to therapy for heart failure patients, it is crucial to remember that the treatment plan is tailored to each individual's needs and must be consistently followed over several months. Throughout this process, adherence is computed cumulatively, allowing us to gain valuable insights into the patient's journey regarding drug adherence. So, it is reasonable to assume that the latent process follows a first-order Markov chain with a certain number of states, called latent states.
By identifying and understanding these latent states, we can reconstruct the path a patient takes regarding the intake of prescribed therapies.

\subsection{Latent Markov model definition}
\label{3.2}
In this work, we propose latent Markov models in a discrete and multivariate scenario, as our primary objective is to observe the measured adherence levels across three distinct classes of drugs.
Let $J$ be the vector of the categorical response variables measured at each time occasion $t=1,..., T$. Denote $Y_{ij}^{(t)}$ the response variable $j \in {1,...,J}$ for the subject $i=1,...,n$ at time $t$, with set of categories $C_j$ with $c_j$ levels coded from $0$ to $c_j-1$. Take 
$\textbf{\textit{Y}}_i^{(t)} = \left(Y_{i1}^{(t)},...,Y_{iJ}^{(t)}\right)$ as the observed multivariate response vector for the subject $i$ at time $t$ and $\widetilde{\textbf{\textit{Y}}_i} = \left(\textbf{\textit{Y}}_i^{(1)},...,\textbf{\textit{Y}}_i^{(T)}\right)$ the relative complete response vector.
Denote with $\widetilde{\textbf{\textit{X}}_i} = \left(\textbf{\textit{X}}_i^{(1)},...,\textbf{\textit{X}}_i^{(T)}\right)$ the complete vector of all individual covariates for the subject $i$.
The model is determined by the observed process $\widetilde{\textbf{\textit{Y}}}_i$ depending on the latent process $\textbf{\textit{U}}_i=\left(U_i^{(1)},...,U_i^{(T)}\right)$. The latent process follows a first-order Markov chain with state space $\{1,...,k\}$, where $k$ identifies the number of latent states.
LM models usually assume that the response vectors $\textbf{\textit{Y}}_i^{(t)} = \left(Y_{i1}^{(t)},...,Y_{iJ}^{(t)}\right)$
are conditionally independent given the latent process $\textbf{\textit{U}}_i$ (\textit{local independence of the response vectors}) and that the elements $Y_{ij}^{(t)}$ are conditionally independent given $U_i^{(t)}$ (\textit{conditional independence of elements}).

The Latent Markov model can be seen as composed of two parts: the measurement model, which describes the conditional distribution of the response variables given the latent variables, and the latent model, which describes the distribution of the latent process.
Individual covariates can be related both to the latent and the measurement model~\citep{cite_34}. In this work, we consider only covariates affecting the initial and transition probability of the latent process.
In particular, the LM model is characterized by three sets of parameters.
\begin{itemize}
    \item The conditional response probabilities $\phi_{jy|u}^{(t)}$ is the probability of observing a response variable $y$ for variable $j$ at time $t$, given the latent state $u\in{1, ..., k}$:
    \begin{equation}
        \text{P}\left(Y^{(t)}_{ij}=y \big| U_i^{(t)}=u\right)= \phi_{jy|u}^{(t)} \qquad j=1,...,J \quad y=0,...,c_j-1.
    \end{equation}
   
    \item The initial state prevalence $\delta_{u|\textbf{\textit{x}}_i^{(1)}}$ is the probability of being in latent state $u \in {1,...,k}$, given the vector of covariates $\textbf{\textit{x}}_i^{(1)}$ for individual $i$:
    \begin{equation}
        \text{P}\left(U_i^{(1)}=u\big|\textbf{\textit{X}}_i^{(1)}=\textbf{\textit{x}}_i^{(1)}\right)= \delta_{u|\textbf{\textit{x}}_i^{(1)}}.
    \end{equation}
    
    \item The transition probabilities $\tau_{u|\Bar{u}\mathbf{x}_i^{(t)}}^{(t)}$, represent the probability of a transition $u$ at time $t$, conditional on being in $\Bar{u}$ at time $t-1$, given the individual vector of covariates $\textbf{\textit{x}}_i^{(t)}$:
    \begin{equation}
        \text{P}\left(U_i^{(t)}=u\big| U_i^{(t-1)}=\Bar{u},\textbf{\textit{X}}_i^{(t)} = \textbf{\textit{x}}_i^{(t)}\right) = \tau_{u|\Bar{u}\textbf{\textit{x}}_i^{(t)}}^{(t)} \qquad t=2,...,T \quad u,\Bar{u}=1,...,k.
    \end{equation}
\end{itemize}

When working with a latent Markov model, it is necessary to compute the marginal distribution to accurately estimate the likelihood of the observed data. 
This distribution, also called \textit{manifest distribution},  is given by: 
\begin{equation}
\begin{split}
    \text{P}(\widetilde{\textbf{\textit{y}}}_i\big|\widetilde{\textbf{\textit{x}}}_i) &
    = \text{P}\left(\widetilde{\textbf{\textit{Y}}}_i=\widetilde{\textbf{\textit{y}}}_i\big| \widetilde{\textbf{\textit{X}}}_i=\widetilde{\textbf{\textit{x}}}_i\right)=\\
    & = \sum_{\textbf{\textit{u}}}\text{P}\left(\widetilde{\textbf{\textit{Y}}_i}=\widetilde{\textbf{\textit{y}}_i}\big|\widetilde{\textbf{\textit{X}}_i}=\widetilde{\textbf{\textit{x}}_i},\textbf{\textit{U}}_i=\textbf{\textit{u}}\right) \times \text{P}\left(\textbf{\textit{U}}_i=\textbf{\textit{u}}\big|\widetilde{\textbf{\textit{X}}_i}=\widetilde{\textbf{\textit{x}}_i}\right)= \\
    &
    = \sum_{\textbf{\textit{u}}}\text{P} \left(\textbf{\textit{U}}_i=\textbf{\textit{u}}|\widetilde{\textbf{\textit{X}}_i}=\widetilde{\textbf{\textit{x}}_i}\right) \times
    \text{P} \left(\widetilde{\textbf{\textit{Y}}_i}=\widetilde{\textbf{\textit{y}}_i}\big|\textbf{\textit{U}}_i=\textbf{\textit{u}}\right)= \\
    &
    = \sum_{\textbf{\textit{u}}} \delta_{u^{(1)}|\textbf{\textit{x}}_i^{(1)}} 
    \prod_{t=2}^{T} \tau_{u^{(t)}|u^{(t-1)}\textbf{\textit{x}}_i^{(t)}}^{(t)} \times \prod_{t=1}^T \prod_{j=1}^J \phi_{jy_{ij}^{(t)}|u^{(t)}}
\end{split}
\end{equation}
The estimation of the parameters is carried out by employing the Expectation–Maximization~\citep{cite_35a} algorithm through the maximization of the log-likelihood for a sample on $n$ independent units~\citep{cite_35}, i.e., $\ell(\boldsymbol{\theta})=\sum_{i=1}^n \log P(\widetilde{\textbf{\textit{y}}}_i \big|\widetilde{\textbf{\textit{x}}}_i)$.
The algorithm’s initialization was performed by applying both deterministic and random initialization techniques to ensure convergence to the global maximum. The number of latent states 
$k$ was determined based on the elbow point of the Bayesian Information Criterion (BIC) curve, which is a commonly used method for model selection in latent variable modeling, and the interpretability of the latent states. This approach is supported by prior literature that highlights the importance of balancing goodness-of-fit and model complexity in selecting the optimal number of states~\citep{cite_interpretability, cite_fmx}.

\subsection{Latent behavioural profiles definition}
\label{3.3}
Once the model has been estimated, a local decoding procedure is performed to obtain a path prediction of the latent states for each sample unit based on the data observed for this unit~\citep{cite_viterbi}.
This procedure is applied to the data to obtain more information on the entire latent process, which associates each patient with the longitudinal path of the states.
For each patient-specific observed data ($\widetilde{\textbf{\textit{x}}}_i$, $\widetilde{\textbf{\textit{y}}}_i$), the Expectation-Maximization algorithm provides the \textit{posterior probabilities} of the latent state $U_i^{(t)}$:
\begin{equation}
    p_{iu}^{(t)} = \text{P}\left(U_i^{(t)}=u \big|\widetilde{\textbf{\textit{Y}}}_i = \widetilde{\textbf{\textit{y}}}_i, \widetilde{\textbf{\textit{X}}}_i = \widetilde{\textbf{\textit{x}}}_i\right) \qquad t=1,...,T \quad u \in {1,...,k}.
\end{equation}
So, the \textit{longitudinal probability profile} for the $i$-th patient, is defined as:
\begin{equation}
    \mathbf{p}_{iu} = \left\{p_{iu}^{(t)} =  P \left(U_i^{(t)}=u \big|\widetilde{\textbf{\textit{Y}}}_i = \widetilde{\textbf{\textit{y}}}_i, \widetilde{\textbf{\textit{X}}}_i = \widetilde{\textbf{\textit{x}}}_i\right), \quad t=1,...,T \right\}.
\end{equation}

By reconstructing the latent process for each patient, we can describe how adherence to therapy varies over months. 
Considering that the latent process comprises $k$ states observed across six periods, this yields an extensive range of potential patterns made by $k^{6}$ distinct combinations of states. To fulfil our objective, we categorized all patterns based on the observed growth or decline trends over time in the states representing the levels of willingness to take polytherapy. This approach allowed us to establish latent-behavioural profiles that synthesize this information over time.

\subsection{Post-hoc analysis}
\label{3.4}
Once the latent profiles were defined, we assessed their impact on patients’ outcomes by applying the Cox regression model to study time to heart failure rehospitalization. This post-hoc analysis represents a critical step in linking the dynamic adherence behaviors, as captured by the latent profiles, to clinically meaningful outcomes. By leveraging the temporal patterns identified in the latent Markov model, this approach enables a comprehensive evaluation of how distinct adherence trajectories influence the risk of rehospitalization.
Figure 1 provides a schematic representation of the entire procedure. In the first step (Step 1, upper panel), dynamic adherence to each drug is assessed monthly over six months of the observation period, allowing us to capture variations in adherence over time. In the second step (Step 2, central panel), latent states are estimated, classifying patients’ adherence behaviors at each time point. These latent states are subsequently used to define latent-behavioral profiles for each patient (Step 3, bottom panel - left), summarizing their adherence patterns across the observation period.
Finally, in Step 4 (bottom panel - right), the association between these latent profiles and survival outcomes is evaluated using Cox regression models, with time to heart failure rehospitalization as the primary endpoint. The Cox model provides estimates of hazard ratios for each latent profile, quantifying the relative risk of rehospitalization associated with distinct adherence patterns. Importantly, the model accounts for key covariates such as age, sex, clinical status and hospitalization during the observation period, aiming to estimate as closely as possible the independent contribution of adherence behaviors.

\begin{figure}[tbp]
\begin{center}
\centerline{\includegraphics[width=0.8\textwidth]{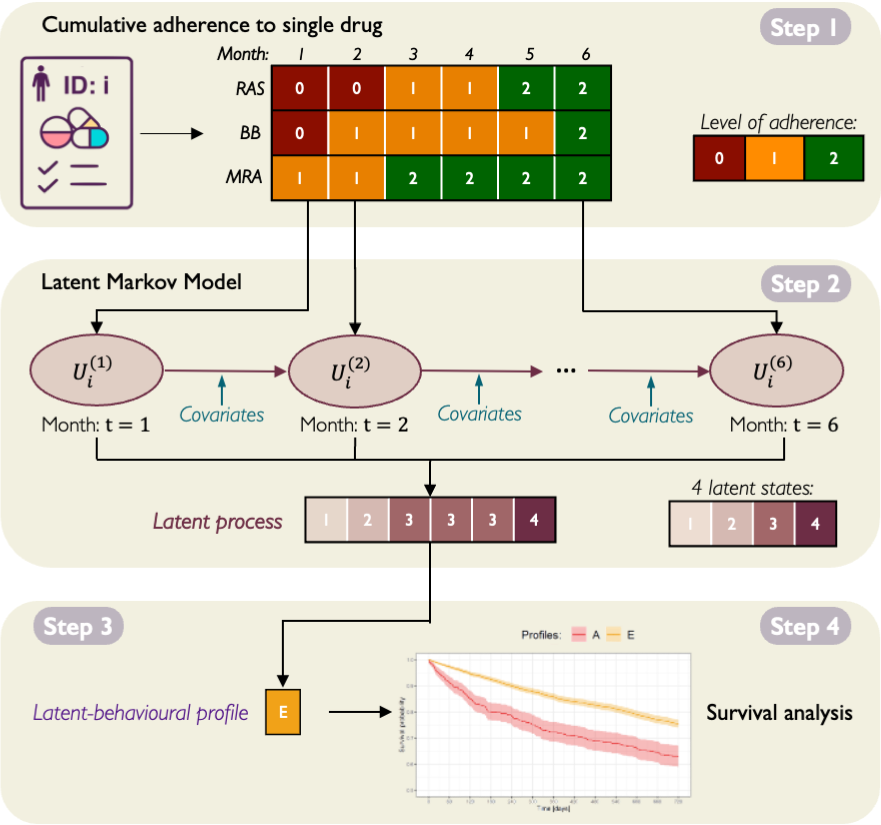}}
\caption{Schematic representation of the methodology used to develop a measure of adherence to polytherapy over time and study its association with patients' survival.}
\label{fig:path_diagram}
\end{center}     
\end{figure}

\subsection{Sensitivity analysis}
Sensitivity analyses were conducted to assess the robustness of the findings under different assumptions. First, the observation period, initially defined as six months, was extended to one year to evaluate whether a longer time frame influenced adherence measurements and subsequent outcomes. Secondly, the follow-up period, originally set at two years, was shortened to one and a half years to test the stability of results under a modified follow-up definition. A graphical representation of these sensitivity analyses can be found in Figure S2. Then, to account for the potential impact of the COVID-19 pandemic, patients with at least one hospitalization for COVID-19 during the observation period were excluded. This adjustment aimed to minimize potential biases introduced by COVID-19-related hospitalizations. 
Finally, to further explore the impact of adherence categorization on the results, the thresholds for adherence, which were initially based on literature-defined cutoffs, were recalibrated using data-driven approaches. This allowed for a comparison of the observed associations under alternative adherence definitions.
These sensitivity analyses aim to reinforce the robustness and validity of the results across varying assumptions and scenarios.

\section{Results}\label{CH4}
In this section, are shown the results obtained by applying the methodologies described in Section~\ref{CH3} to the dataset described in Section~\ref{CH2}. All the analyses were carried out using the free software {\tt{R}}~\citep{cite_31}, in particular {\tt{LMest}}~\citep{cite_32} package. 

\subsection{Latent Markov Model estimation}\label{4.3}
For each month $t =1,...,6$ of the observation period, let $\textit{J} = \{RAS, BB, MRA\}$ be the set of drugs considered, we denote with $Y_{ij}^{(t)}$ the level of adherence to the drug $j$ for each patient $i$. The relative sets of response categories identified were coded as follows: $C_j = \{0: low, 1: middle, 2: high\}$ adherence to the drug $j$.
The first step for obtaining the final multivariate latent markov model (LM) involves selecting the covariates to be included in the latent model. This involves incorporating three time-fixed covariates (age, sex, and hospitalization during the observation period) and one time-varying covariate (MCS). The second step aims to identify the number of latent states $k$ of the latent process. 
Multivariate LM models were fitted with all covariates in the latent model, increasing $k$ from 1 to 5.  The optimal number of latent states $k$ was identified based on the elbow point of the Bayesian Information Criterion curve and the interpretability of the latent states (Figure S3). Recalling that, we are interested in classifying patients based on their adherence levels to polytherapy and then choosing a $k$ capable of synthesizing the general behaviour over the three classes of drugs considered. The selected number of latent states is $k=4$, identifying a latent process with four levels of adherence. In the final model, the parameter were defined as:
\begin{itemize}
    \item Initial probabilities were associated with the patient’s \textit{age}, \textit{sex}, \textit{hosp} and \textit{MCS}, and for each patient $i$ are defined as:
    \begin{equation}
        \log \frac{\delta_{u|age, sex, hosp, MCS}}{\delta_{1|age, sex, hosp, MCS}} = \beta_{0u}+\beta_{1u} \cdot age_i +
        \beta_{2u} \cdot sex_i+ \beta_{3u} \cdot hosp_i+
        \beta_{4u} \cdot MCS_i
        \label{eq10}
    \end{equation}
    \item Transition probabilities were associated with the patient’s \textit{age}, \textit{sex}, \textit{hosp} and \textit{MCS}, and for each patient $i$ are defined as:
    \begin{equation}
        \log \frac{\tau^{(t)}_{u|\bar{u}(age, sex, hosp, MCS)}}{\tau^{(t)}_{\bar{u}|\bar{u}(age, sex, hosp, MCS)}} = \gamma_{0\bar{u}u}+\gamma_{1\bar{u}u} \cdot age_i +
        \gamma_{2\bar{u}u} \cdot sex_i+ \gamma_{3\bar{u}u} \cdot hosp_i+
        \gamma_{4\bar{u}u} \cdot MCS^{(t)}_i
        \label{eq11}
    \end{equation}
\end{itemize}

\begin{figure}[tbp]
\begin{center}
\centerline{\includegraphics[width=1\textwidth]{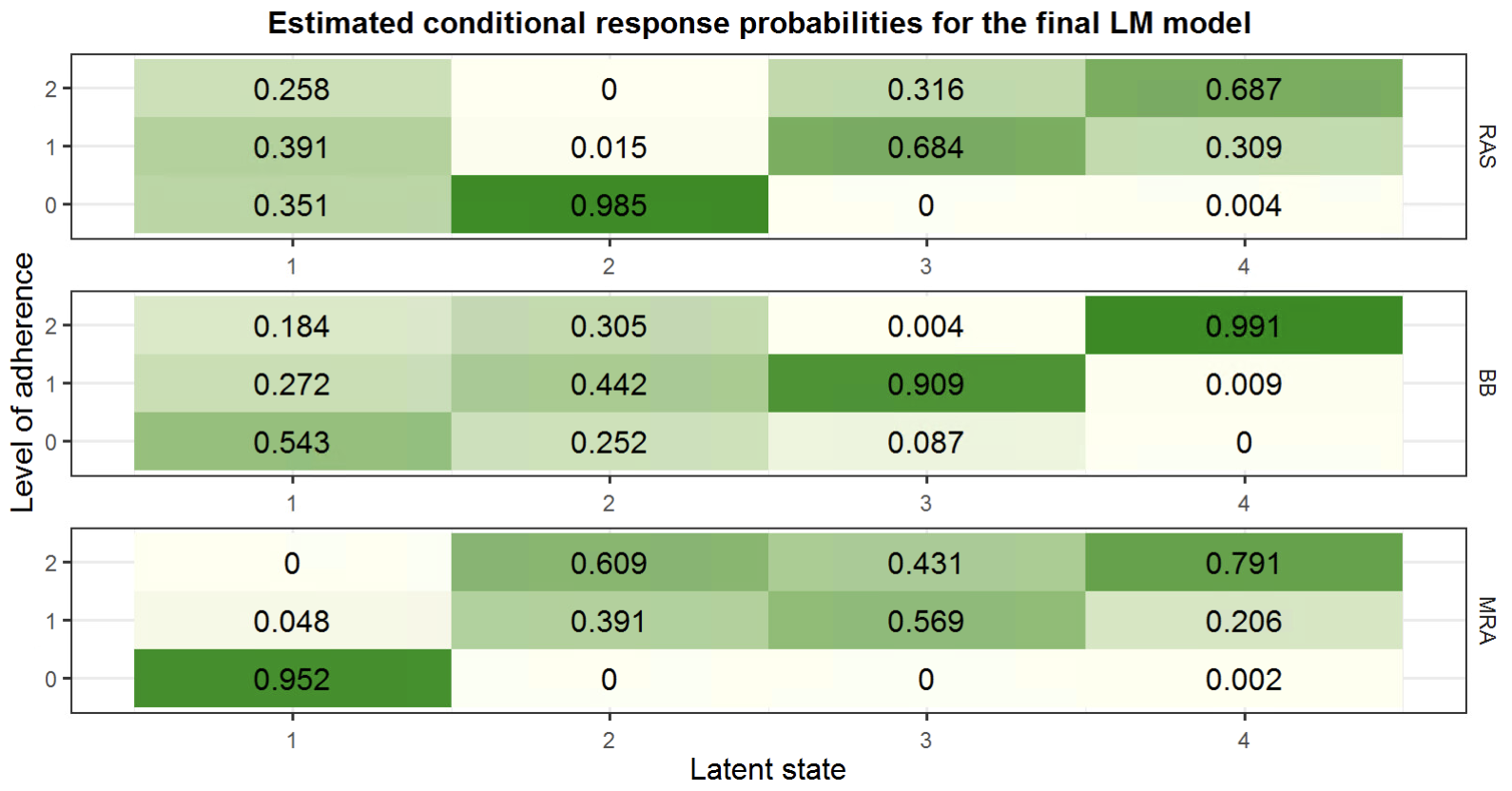}}
    \caption{Estimated conditional response probabilities $\phi_{jy|u}$ for the final multivariate LM model. Each panel refers to different classes of drugs $j \in \textit{J} = \{RAS, BB, AA\}$. Each row refers to a response category $C_j \in \{0, 1, 2\}$. Each column refers to a latent state $u = \{1, 2, 3, 4\}$.}
    \label{fig:matrix_cond_prob}
\end{center}    
\end{figure}

Figure~\ref{fig:matrix_cond_prob} shows the estimated conditional response probabilities $\phi_{jy|u}$ for each drug $j$ under the selected model. These probabilities can be helpful in interpreting the latent states.  
Since, in our model, the latent process represents an overall measure of adherence not directly observable/measurable, we indicate this process as the level of \textit{willingness} of the patients to take the prescribed drugs (see~\ref{3.1} for the definition). Based on this interpretation, the following latent state labelling was derived: State 1 middle-low willingness to be adherent to polytherapy; State 2 middle willingness to be adherent to BB and MRA and low to RAS; State 3: middle-strong willingness to be adherent and finally, State 4; strong willingness to be adherent to polytherapy.
It is crucial to keep in mind that if a patient does not purchase a specific drug in the entire observation period, we hypothesized that he does not have the prescription, so he will be categorized as a non-user and adherence to that drug will not be assessed. This approach guarantees that the evaluation of patient adherence is exclusively focused on the drugs supposed to be prescribed by clinicians, preventing any unwarranted penalization in the assessment of overall adherence. For example, if a patient $i$ never purchases MRA but takes RAS and BB constantly during the observation period, he will be considered a user of them and a non-user of MRA, and the model will assign him to the latent state 4 for each month.

Table~\ref{tab:est_regre_param} displays the estimated regression parameters influencing the logit for the initial probabilities $\hat{\boldsymbol{\beta}} = (\hat{\beta}_{1u}, \hat{\beta}_{2u}, \hat{\beta}_{3u}, \hat{\beta}_{4u})$.
Focusing on age, the estimate is negative for State 4, indicating that younger individuals were more likely to report higher drug adherence during the first month compared to older ones.
For sex, the estimates are not statistically significant across all states, suggesting that sex does not substantially influence adherence behavior in this context.
Regarding the MCS, all estimates are negative, indicating that patients with better health conditions (lower comorbidity burden) were more likely to adhere to their prescribed drugs in the first month compared to those in worse clinical conditions.
Finally, the estimates for hospitalization during the observation period are negative for all states compared to State 1, indicating that patients who were hospitalized during the observation period were less likely to exhibit high adherence to their prescribed drugs in the first month.

\begin{table}[h]
\caption{Estimated regression parameters affecting the distribution of the initial probabilities in Equation~(\ref{eq10}).}\label{tab:est_regre_param}
\begin{tabular}{lclll}
\toprule
\multicolumn{5}{c}{\textbf{Regression parameters for initial probabilities}} \\
 & \textit{\textbf{u}} & 2 & 3 & 4 \\ 
 \hline
\textbf{Age} & $\hat{\beta}_{1u}$ & 0.0069 & -0.0049 & -0.0100*** \\
\textbf{Gender Male} & $\hat{\beta}_{2u}$ & -0.0645 & -0.0383 & -0.0800 \\ 
\textbf{MCS} & $\hat{\beta}_{3u}$ & -0.0044 & -0.0164*** & -0.0125*** \\ 
\textbf{Hospitalization$^\dagger$} & $\hat{\beta}_{4u}$ & -0.5152*** & -0.9098*** & -0.7741***\\ 
\botrule
\end{tabular}
\text{Significance: *(\%10), **(\%5), ***(\%1).}\\
\text{$^\dagger$ Hospitalization during the observation period.}
\end{table}

Table~\ref{tab:est_tran_rate} provides the mean estimated transition probabilities $\hat{\tau}_{u|\bar{u}}$ between latent states, reflecting the likelihood of patients transitioning between adherence states over time.
The results indicate that States 3 and 4, representing middle-strong and strong willingness to adhere, are the most stable, with high probabilities of remaining in the same state (87.9\% and 94.3\%, respectively). This stability suggests that patients who achieve higher adherence levels are likely to maintain them over time.
In contrast, States 1 and 2, associated with middle-low and low RAS adherence, exhibit greater variability and a higher likelihood of transitioning, particularly to State 3. Moreover, the transition probabilities reveal an overall trend of increasing adherence states (e.g., transitioning to higher latent states) compared to decreasing values.

\begin{table}[h]
\caption{Mean estimates of the transition probabilities between latent states in Equation~(\ref{eq11}).}\label{tab:est_tran_rate}
\begin{tabular}{ccccc}
\toprule
\multicolumn{5}{c}{\textbf{Transition probabilities from $\bar{u}$ to $u$ $(\hat{\tau}_{u|\bar{u}})$}} \\ 
\hline
\textbf{$\bar{u}$ / $u$} & 1 & 2 & 3 & 4\\
\hline
1 & 0.697 & 0.038 & 0.245 & 0.019\\
2 & 0 & 0.819 & 0.125 & 0.055\\
3 & 0.004 & 0.009 & 0.879 & 0.107\\ 
4 & 0 & 0.004 & 0.053 & 0.943\\ 
\botrule
\end{tabular}
\end{table}

\subsection{Latent-behavioural profiles}\label{4.4}
Once the model parameters are estimated, the longitudinal probabilities profiles $p_{iu}$ are reconstructed. For each patient $i$, using local decoding, we predict the longitudinal profile in the observation period, i.e., the sequence of latent states visited during the observation period. Looking at the temporal order of the latent states in each latent process, seven latent-behavioural profiles may be exploited.
\begin{itemize}
    \item Profile A: patients always middle-low willingness to be adherent to polytherapy
    \item Profile B: patients always middle willingness to be adherent to BB and MRA and low to RAS 
    \item Profile C: patients always middle-strong willingness to be adherent to polytherapy 
    \item Profile D: patients always strong willingness to be adherent to polytherapy 
    \item Profile E: patients with increasing levels of willingness to be adherent 
    \item Profile F: patients with decreasing levels of willingness to be adherent 
    \item Profile G: patients with varying willingness to be adherent 
\end{itemize}

Looking at the profiles, we noticed a balance with respect to demographic and clinical characteristics (age, sex, and MCS) demonstrating minimal variation between groups, as evidenced in Table S3  by standardized mean differences consistently below 0.1.
This balanced distribution strengthens the reliability of our comparisons between profiles and suggests that any observed differences in survival outcomes can be more confidently attributed to the profile effects rather than underlying demographic or clinical disparities.
 
\subsection{Survival analysis}\label{4.5}
Table~\ref{tab:hr} displays the impact of different profiles on time to heart failure rehospitalization for the fitted Cox's regression model adjusted for age, sex and multisource comorbidity score. All the profiles are statistically different from reference profile A, identifying those patients who are always middle-low adherent to the therapy. Being in the other profile is a protective factor for the survival probability; in particular, being in profile C decreases by 44\% [95\% CI, 30\%-55\%] the likelihood of rehospitalization for HF while being in D by 56\% [95\%CI, 47\%-64\%].

\begin{figure}[tbp]
\begin{center}   
\centerline{\includegraphics[width=0.8\textwidth]{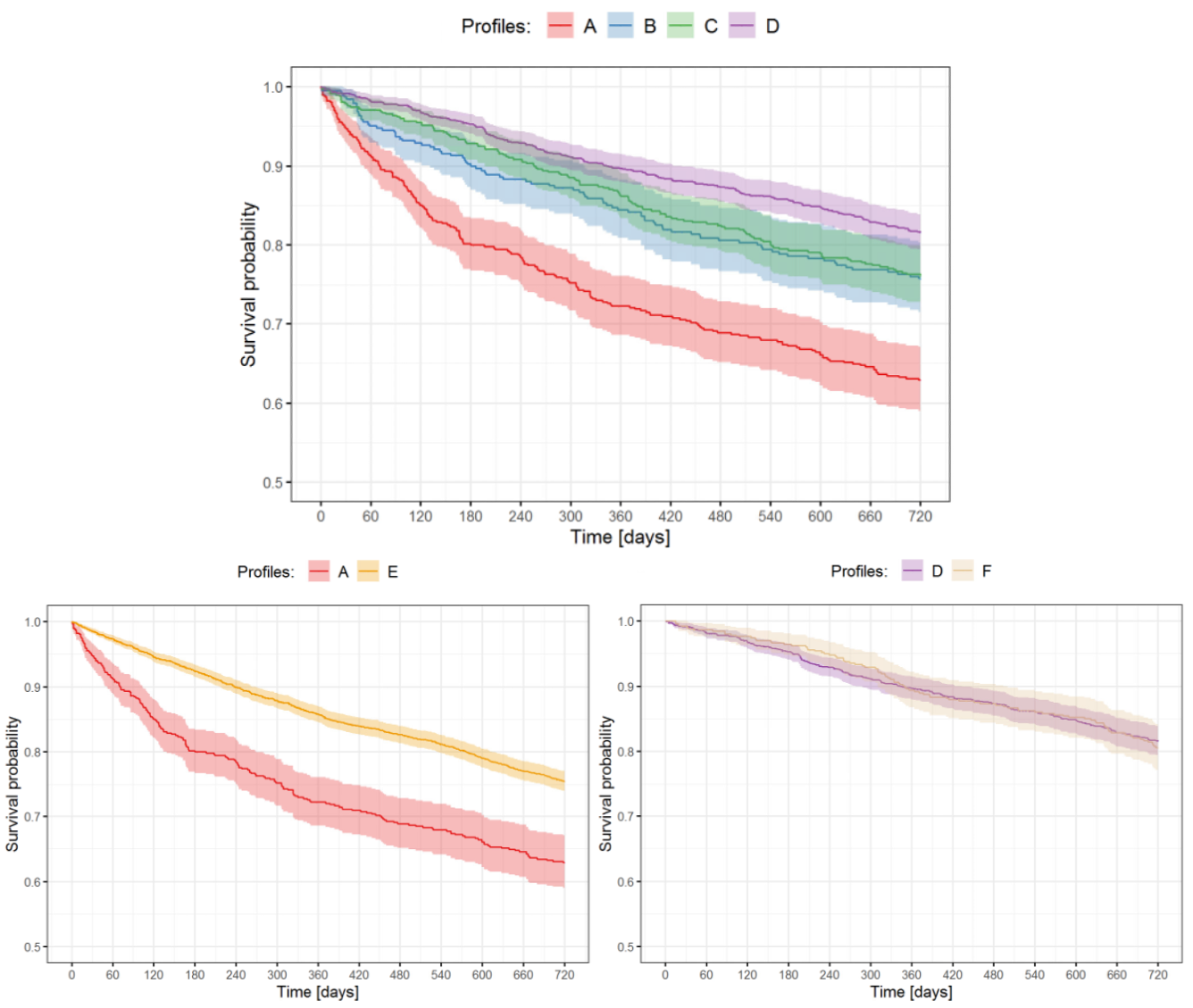}}
\caption{KM survival curves for the time to heart failure re-hospitalization stratified by latent-behavioural profile. Top: \{A, B, C, D\}; Bottom left: \{A, E\}; Bottom right: \{D, F\}.}
\label{fig:surv_all}
\end{center}
\end{figure}

\begin{table}[h]
\caption{Adjusted cause-specific hazard ratios (95\% confidence interval) and p values of Cox's model.}\label{tab:hr}
\begin{tabular}{lll}
\toprule
Latent-behavioural profiles & HR (95\% CI) & p value \\ \hline
A (n = 569) & reference &  \\
B (n = 388) & 0.60 (0.47 - 0.77) & \textbf{$<$0.001} \\
C (n = 584) & 0.56 (0.45 - 0.70) & \textbf{$<$0.001} \\
D (n = 1212) & 0.44 (0.36 - 0.53) & \textbf{$<$0.001} \\
E (n = 3259) & 0.58 (0.50 - 0.68) & \textbf{$<$0.001} \\
F (n = 513) & 0.43 (0.34 - 0.55) & \textbf{$<$0.001} \\
G (n = 293) & 0.57 (0.43 - 0.75) & \textbf{$<$0.001} \\
\botrule
\end{tabular}
\end{table}
We now exploit the comparison of these results by estimating the Kaplan-Meier survival curves stratified by patient profiles. In Figure~\ref{fig:surv_all} (top), a difference in the curves is observed when comparing patients with the same latent state throughout the observation period. High-adherent patients have a greater probability of survival than those with middle and low adherence. Patients in groups B and C exhibit similar survival curves that align with the respective levels of adherence.
The estimate showed that patients who adhered to their prescribed therapies survived 3.66 [95\% CI, 2.85-4.46] months longer, on average, than non-adherence when following up patients for two years. 
In Figure~\ref{fig:surv_all} (bottom left), we can observe that the survival probabilities of consistently middle-low adherent patients are lower than those who increase their adherence over time: moving from being low-adherent to high-adherent (patients in E) significantly increases these probabilities. Figure~\ref{fig:surv_all} (bottom right) reveals comparable outcomes between consistently high-adherent patients and those exhibiting declining adherence levels. This can be explained by the fact that in group G, adherence levels decrease over time, but most patients begin with a latent State of 3 or 4. As a result, they do not have enough time to transition to low adherent status. Only three patients start with a latent State 2 and decrease to State 1.

\subsection{Sensitivity analysis}\label{4.6}
In our sensitivity analyses, we assessed the robustness of our findings under different conditions. 
When evaluating adherence over a one-year follow-up period, the cohort was reduced to 1,641 patients with at least one year of follow-up. The latent process characterizations (Figure S4) and hazard ratios from the Cox model closely reflected the results of the main analysis, though with reduced statistical power.
Shortening the follow-up period to 1.5 years resulted in hazard ratios that were comparable to those of the primary analysis.
When limiting the analysis to patients who had not been hospitalized for COVID-19 during the observation period, both the conditional expectation matrices for the latent states (Figure S5) and the hazard ratios were consistent with those of the main analysis.
In the final sensitivity analysis, we modified the adherence categorization thresholds to better reflect the observed distribution of adherence, rather than relying on literature-based cutoffs. As a result, Equation~(\ref{eq1.1}) was revised to:
\begin{equation}
         \label{eq1.2}
        \text{adherence}_i^{(t)} = \begin{cases}
            0, &\text{if } \frac{\text{com\_month}_i(t)}{t} \in \text{[0, 0.35)},\\
            1, &\text{if } \frac{\text{com\_month}_i(t)}{t} \in \text{[0.35, 0.9),}\\
            2, &\text{if } \frac{\text{com\_month}_i(t)}{t} \in \text{[0.9, 1].}\\
            \end{cases}
\end{equation}
Using this formula to define drug adherence yielded similar results from the latent Markov model (Figure S6) and comparable hazard ratios in the Cox regression model. 
A comparison of the adjusted hazard ratios (95\% CI) from the Cox regression models across all sensitivity analyses is presented in (Table S4). These consistent findings across multiple sensitivity analyses strengthen the reliability of our primary results and suggest that our conclusions are robust to variations in both follow-up duration and adherence categorization methods.

\section{Discussion and Conclusions}\label{CH5}
We proposed an innovative methodological pipeline that considers the dynamic nature of drug adherence, which is often overlooked in traditional research. The key point of our proposal is the ability to assess both the dynamic and multivariate dimensions simultaneously. Considering the ever-evolving landscape of prescription patterns in heart failure settings, where novel combinations of medications are continually prescribed~\citep{cite_40}, our approach acknowledges the intrinsic complexity of treatment regimens. This work contributes to a more holistic comprehension of the relationship between treatment strategies and patient survival, highlighting the importance of personalized and adaptive interventions in the ever-evolving field of medical care.
Moreover, the necessity to describe complex patterns of care over time without losing the heterogeneity and variability of this information, coupled with the awareness that we may only have a partial view of the phenomenon, rise to the idea of applying the Latent Markov model to study adherence to drugs for heart failure patients. This model assumes the existence of a latent process where both its initial and transition probabilities depend on a set of individual covariates. Since this process explains a characteristic of the observed single-adherence to drugs over time, which is not directly observable, we interpret it as the patient's willingness to take the purchased therapies. This approach ensures that individuals who do not use specific drugs are not penalized in their evaluation of adherence willingness, avoiding the restriction to ``users-only" cohorts ~\citep{cite_42,cite_43}.

Importantly, our contribution does not lie solely in the application of LMMs, which are well-established in the field, but rather in the innovative combination of them with dynamic adherence modeling for polytherapy in heart failure patients. Moreover, we integrated the results of these models into prognostic frameworks, such as Cox regressions, to assess the impact of adherence behaviors on clinical outcomes, particularly time to heart failure rehospitalization. This comprehensive pipeline represents a novel contribution to the study of adherence, as it simultaneously considers temporal and multivariate dimensions while linking them to survival analysis.

The obtained latent process is composed of four latent states, representing the different levels of patients' willingness to adhere to prescribed therapies. This approach not only allows us to capture dynamic adherence patterns but also provides a robust framework for identifying clinically relevant behavioral profiles. 

Examining the estimated regression parameters affecting the distribution of initial probabilities,  we noticed that younger patients exhibited better adherence early on, suggesting that older individuals may face greater barriers to maintaining their regimens, such as cognitive or logistical challenges. 
Patients with higher comorbidity burdens (higher MCS) were less adherent, underscoring the complexity of managing multiple conditions simultaneously. 
Hospitalization during the observation period was associated with lower adherence. This suggests that the disruption caused by hospitalization negatively affects medication routines, emphasizing the need for robust post-discharge interventions to re-establish adherence behaviors. Together, these findings underscore the value of personalized and dynamic approaches to monitoring and enhancing adherence, particularly in vulnerable subgroups, to optimize therapeutic outcomes and reduce adverse events.
On the other hand, the stability observed in patients with higher adherence levels (States 3 and 4) suggests that once strong adherence is established, it tends to persist. This finding underscores the importance of sustaining high adherence through consistent follow-up and support, as these patients are less likely to relapse into lower adherence states.
Conversely, the higher likelihood of transitioning from low adherence states (States 1 and 2) to middle-strong adherence (State 3) indicates that improving adherence among less adherent patients is achievable. This highlights the potential for targeted interventions to shift patients toward more stable and beneficial adherence behaviors, emphasizing the importance of early and individualized adherence strategies.
The overall trend of progression toward higher adherence states also reflects the dynamic and evolving nature of adherence behavior. This improvement over time may result from increased patient engagement, medication adjustments, or reinforcement through clinical interactions. These findings reinforce the necessity for adaptive approaches that account for temporal changes in adherence, tailoring strategies to the evolving needs of patients.

By employing the LM model, we can define patient-specific latent processes that keep track of the dynamic evolution of adherence willingness over months. These processes revealed clinically significant patterns, allowing us to profile patients based on whether their adherence levels remain consistent, increase, or decrease over time. The results from the Cox regression model and Kaplan-Meier survival curves provide robust evidence for the association between adherence profiles and clinical outcomes. Patients with consistently high adherence (Profiles C and D) demonstrated a significantly lower risk of heart failure rehospitalization compared to the reference group (Profile A), with reductions of 44\% and 56\%, respectively. These findings underscore the protective effect of strong adherence to therapy and the importance of maintaining it over time. Moreover, the Kaplan-Meier survival curves further elucidate the benefits of improving adherence. Patients in Profile E, who transitioned from low to high adherence, exhibited markedly improved survival probabilities compared to those with persistently low adherence. This suggests that even patients starting with suboptimal adherence can achieve better outcomes through timely interventions that enhance their adherence behaviors.
Interestingly, Profile F, characterized by declining adherence, exhibited survival probabilities comparable to consistently high-adherent patients. This can be attributed to the fact that most individuals in this group began with high adherence (States 3 or 4) and did not have sufficient time to transition to lower adherence states. This finding highlights the importance of early adherence monitoring and intervention to prevent declines.
These results collectively emphasize the dynamic nature of adherence and its substantial impact on clinical outcomes. 

The sensitivity analyses confirmed that these findings are robust across different definitions of observation and follow-up periods, as well as alternative adherence threshold categorizations. Moreover, excluding COVID-19-related hospitalizations did not alter the main results, indicating that the pandemic did not substantially bias our findings. The association between adherence and improved survival outcomes aligns with prior research, further confirming that consistent adherence to prescribed therapies leads to fewer hospitalizations and better patient prognoses~\citep{cite_5, cite_21}. This is particularly relevant given the added healthcare challenges posed by the COVID-19 pandemic, which disrupted routine care and may have impacted adherence patterns. 

Our study highlights the need to evaluate the persistence or change in adherence over time, as improving individual behavior positively impacts clinical outcomes, while declining adherence can worsen health status. Patients who consistently adhere to their medications benefit from extended survival and lower rehospitalisation rates. Beyond evaluating temporal trends, our methodology also assesses the effects of adherence to different therapies over time, demonstrating that high motivation to follow polytherapy significantly reduces the risk of adverse outcomes.
The developed procedure is a flexible approach that is capable of accommodating different numbers and types of drugs. This adaptability is particularly critical given the introduction of novel pharmacological treatments, such as sodium-glucose co-transporter 2 (SGLT2) inhibitors, into standard protocols~\citep{cite_41}.

Some limitations of the present study should also be acknowledged. Firstly, the limited clinical information included in administrative databases characterizing heart failure does not allow for an accurate assessment of the severity of the disease. However, precisely because we lack this detailed information, latent Markov models are even more useful as they can effectively capture the complex and dynamic nature of data and provide insights into the underlying mechanisms and factors that influence patient outcomes. Secondly, distinguishing between individuals who initiated treatment during the year and those who did not adhere to treatment from the beginning of the year can pose a challenge when using adherence measures based on coverage months. Sensitivity analyses confirmed that adherence categorization thresholds and observation periods did not significantly alter the results, suggesting robustness to this limitation.
Finally, while our methodology provides valuable insights, the choice of latent profiles and state labels remains somewhat subjective. Future research should explore additional validation methods and alternative definitions to enhance reproducibility and generalizability.

While our approach was guided by clinical reasoning to achieve interpretable and clinically meaningful results, it is important to acknowledge the existence of alternative methodological choices. The latent profiles identified in our study offer a valuable foundation for analyzing their associations with future outcomes. However, they should not be regarded as the only valid approach. Advancements in methodological frameworks, such as the hidden Markov model proposed by ~\citep{Pandolfi2023}, could enhance the robustness of our findings by addressing complex missing data patterns, including partially missing outcomes, intermittent missingness, and informative dropout.
Furthermore, future research could benefit from exploring alternative analytical approaches, such as state sequence analysis~\citep{cite_ssa}, to complement and expand upon the insights gained from our study. By leveraging these innovative methodologies, it may be possible to deepen our understanding of adherence behaviors and their intricate relationships with clinical outcomes, ultimately contributing to more precise and effective patient management strategies.\\

In conclusion, our approach illustrates the benefits of studying dynamic adherence to polytherapy. The results underscore the significance of regular follow-up and interventions to enhance adherence levels, offering dual benefits by improving patient outcomes and positively impacting the healthcare system through potential cost reductions. Notably, studying polytherapy adherence, as opposed to adherence to single drugs, provides valuable insights into medication management skills and patient behavior.

\textbf{Supplementary information}
An additional file is available.\\
\textbf{Figure S1.} Study flow chart of patient inclusion/exclusion criteria.\\
\textbf{Figure S2.} Graphical representation of the sensitivity analysis.\\
\textbf{Figure S3.} Model selection of the latent state for the Latent Markov Model.\\
\textbf{Figure S4.} Sensitivity 1-year of observation period: estimated conditional response probabilities.\\
\textbf{Figure S5.} Sensitivity non COVID-19: Estimated conditional response probabilities.\\
\textbf{Figure S6.} Sensitivity adherence threshold: estimated conditional response probabilities.\\
\textbf{Table S1.} ICD-10 and ATC codes definition.\\
\textbf{Table S2.} Baseline characteristics of the cohort.\\
\textbf{Table S3.} Patients characteristics stratified by latent behavioural profiles.\\
\textbf{Table S4.} Hazard Ratio of Cox's model for the sensitivity analysis comparison.

\section*{Declarations}
\textbf{Acknowledgment. }The authors acknowledge the support of Regione Lombardia, and in particular Marco Forlani, Barbara Antonelli, and Olivia Leone, for facilitating the data-sharing agreement and access to the administrative healthcare databases used in this research.\\
\textbf{Funding. }No sources of funding were used to conduct this study or prepare this manuscript.\\
\textbf{Conflict of interest/Competing interests. }All the authors have no potential conflicts of interest that might be relevant to the contents of this manuscript.\\
\textbf{Ethics approval and consent to participate. }The study (i) is exempt from patients informed consent (according to General Authorization for the Processing of Personal Data for Scientific Research Purposes Issued by the Italian Privacy Authority on August 10, 2018; https://www.garanteprivacy.it/web/guest/home/docweb/-/docweb-display/docweb/9124510); (ii) provides sufficient guarantees of individual records anonymity; and (iii) was designed according to quality standards of good practice of observational research based on secondary data.\\
\textbf{Data availability. }The data that support the findings of this study are available from the Lombardy Region, but restrictions apply to the availability of these data, which were used under license for the current study, and so are not publicly available. \\
\textbf{Code availability. }All codes and simulated data for reproducing the procedures are available.

\bibliography{sn-bibliography}

\end{document}